\documentclass[%
 reprint,
 amsmath,amssymb,
 aps,
]{revtex4-2}

\usepackage{graphicx}
\usepackage{dcolumn}
\usepackage{bm}
\usepackage{xcolor}
\usepackage{algorithm} 
\usepackage{algpseudocode} 
\usepackage{tabularx}

\begin{document}


\title{Maximum likelihood estimation of burst-merging kernels for bursty time series}

\author{Tibebe Birhanu}
\affiliation{Department of Physics, The Catholic University of Korea, Bucheon, Republic of Korea}

\author{Hang-Hyun Jo}
\email{h2jo@catholic.ac.kr}
\affiliation{Department of Physics, The Catholic University of Korea, Bucheon, Republic of Korea}

\date{\today}

\begin{abstract}
Various time series in natural and social processes have been found to be bursty. Events in the time series rapidly occur within short time periods, forming bursts, which are alternated with long inactive periods. As the timescale defining bursts increases, individual events are sequentially merged to become small bursts and then bigger ones, eventually leading to the single burst containing all events. Such a merging pattern has been depicted by a tree that fully reveals the hierarchical structure of bursts, thus called a burst tree. The burst-tree structure can be simply characterized by a burst-merging kernel that dictates which bursts are merged together as the timescale increases. In this work, we develop the maximum likelihood estimation method of the burst-merging kernel from time series, which is successfully tested against the time series generated using several model kernels. We also apply our method to some empirical time series from various backgrounds. Our method provides a useful tool to precisely characterize the time series data, hence enabling to study their underlying mechanisms more accurately.
\end{abstract}

\maketitle

\section{Introduction}

Complex systems have been characterized by the complex interaction structure among elements of the systems. Such interaction structure can be described in terms of the network where links are formed between interacting elements or nodes~\cite{Barabasi2016Network, Newman2018Networks, Menczer2020First, Dorogovtsev2022Nature, Easley2010Networks, Barrat2008Dynamical}. To study the dynamics of complex systems one can adopt the framework of temporal networks, where links are considered to exist only at the moment of interaction~\cite{Holme2012Temporal, Holme2019Temporal, Masuda2016Guide}. The time series of those interaction events turn out to be inhomogeneous or bursty; events in the time series rapidly occur within short time periods, forming bursts, which are alternated with long inactive periods~\cite{Barabasi2005Origin, Karsai2018Bursty, Jo2023Bursty}. This bursty phenomenon has been observed in various systems, whether being physical~\cite{Bak1987Selforganized, Jensen1998Selforganized, Wheatland1998WaitingTime, Corral2004Longterm}, biological~\cite{Beggs2003Neuronal, Petermann2009Spontaneous, Kemuriyama2010Powerlaw}, or social~\cite{Goh2008Burstiness, Crane2008Robust, Rybski2009Scaling, Karsai2018Bursty, Jo2020Bursttree, Choi2021Individualdriven}. Thus, proper characterization methods of the bursty time series are important to understand the dynamics of complex systems.

A number of methods have been proposed to characterize the bursty behaviors in various empirical time series, including those from temporal networks~\cite{Karsai2018Bursty, Jo2023Bursty}. Among others, the burst-tree decomposition method~\cite{Jo2020Bursttree} provides the comprehensive picture of temporal correlations in the time series by fully revealing the hierarchical structure of bursts over the entire range of timescales. Precisely, for a given timescale $\Delta t$, events in the time series can be clustered into bursts that are separated from each other by time intervals larger than $\Delta t$; it means that time intervals between two consecutive events within the burst are smaller than or equal to $\Delta t$~\cite{Karsai2012Universal}. The number of events in a burst is called a burst size. Thus, as the timescale defining bursts increases, individual events are sequentially merged to form small bursts and then bigger ones, eventually ending up with the single burst containing all events. Such a merging pattern can be depicted by a tree network, which fully reveals the hierarchical structure of bursts, thus it is called a burst tree. By definition of the bursts, the time interval between two consecutive events, or an interevent time (IET), plays an important role in characterizing the temporal correlations. Interestingly, the burst-tree structure derived from various empirical datasets was shown to share similar properties such as heavy-tailed burst size distributions and positive correlations between two consecutive burst sizes over a wide range of timescales, irrespective of the data-specific IET distributions~\cite{Jo2020Bursttree}. It implies that there might be some common mechanisms behind similar burst-tree structure in diverse systems.

The burst-tree structure derived from the data can be seen as a realization of the merging process starting from individual events~\cite{Jo2020Bursttree, Birhanu2025Bursttree}. These events are sequentially merged according to some selection rule or a burst-merging kernel dictating which bursts are merged together as the timescale increases. It is based on the observation that two successive bursts that are separated by the IET larger than a timescale will be merged when the timescale increases to exceed the IET between those bursts. The empirically estimated burst-merging kernels have revealed the preferential and assortative merging behaviors, which are responsible for the heavy-tailed distributions of burst sizes and positive correlations between two consecutive burst sizes. In this sense, the estimation of the burst-merging kernel is important to precisely characterize the time series data as well as to study their underlying mechanisms accurately. The estimation method of the burst-merging kernel proposed in Ref.~\cite{Jo2020Bursttree} is useful but not rigorous, requiring us to devise a more statistically sound method. 

In this work, we develop the maximum likelihood estimation method of the burst-merging kernel, inspired by the estimation method of the attachment kernel in the context of network science~\cite{Pham2015PAFit}; when a network grows with newly added nodes, each new node chooses a subset of existing nodes for making connections to them, following some rule. Such a rule is often implemented in terms of the attachment kernel dictating which nodes are chosen among existing nodes. For example, in the case of the preferential attachment models for growing networks~\cite{Barabasi1999Emergence, Krapivsky2000Connectivity, Barabasi2016Network}, the kernel for an existing node has been given as an increasing function of its degree, meaning that nodes with higher degrees are more likely to be chosen by newly added nodes. It typically leads to the heavy-tailed or power-law degree distributions. Therefore, the attachment kernel is one of the most important factors shaping the structure of growing networks. Several estimation methods of the attachment kernel from empirical network data have been proposed~\cite{Newman2001Clustering, Jeong2003Measuring, Sheridan2012Measuring, Kunegis2013Preferential, Pham2015PAFit}. Similarly, the burst-merging kernel dictating which bursts are chosen to be merged is the central element in understanding the burst-tree structure. After developing the maximum likelihood estimation method of the burst-merging kernel for the time series, we test the validity of our method with the time series generated using several model kernels, including constant, sum, and product kernels~\cite{Birhanu2025Bursttree}. Then, we show that our method is applied to some empirical time series from various backgrounds.

The paper is organized as follows: In Sec.~\ref{sec:burst_tree} we describe the burst-tree decomposition method and the role of burst-merging kernels in the burst-tree structure. In Sec.~\ref{sec:estimation} we develop the maximum likelihood estimation method of the burst-merging kernel for the time series, test its validity using several model kernels, and finally apply the method to some empirical time series. We conclude our paper in Sec.~\ref{sec:conclusion}.

\section{Burst-tree structure}\label{sec:burst_tree}

We first introduce the burst-tree decomposition method for analyzing the time series. We then discuss the crucial role of the burst-merging kernel in understanding the hierarchical structure of bursts in the burst tree.

\subsection{Burst-tree decomposition method}\label{subsec:decompose}

\begin{figure}[!t]
\centering
\includegraphics[width=\columnwidth]{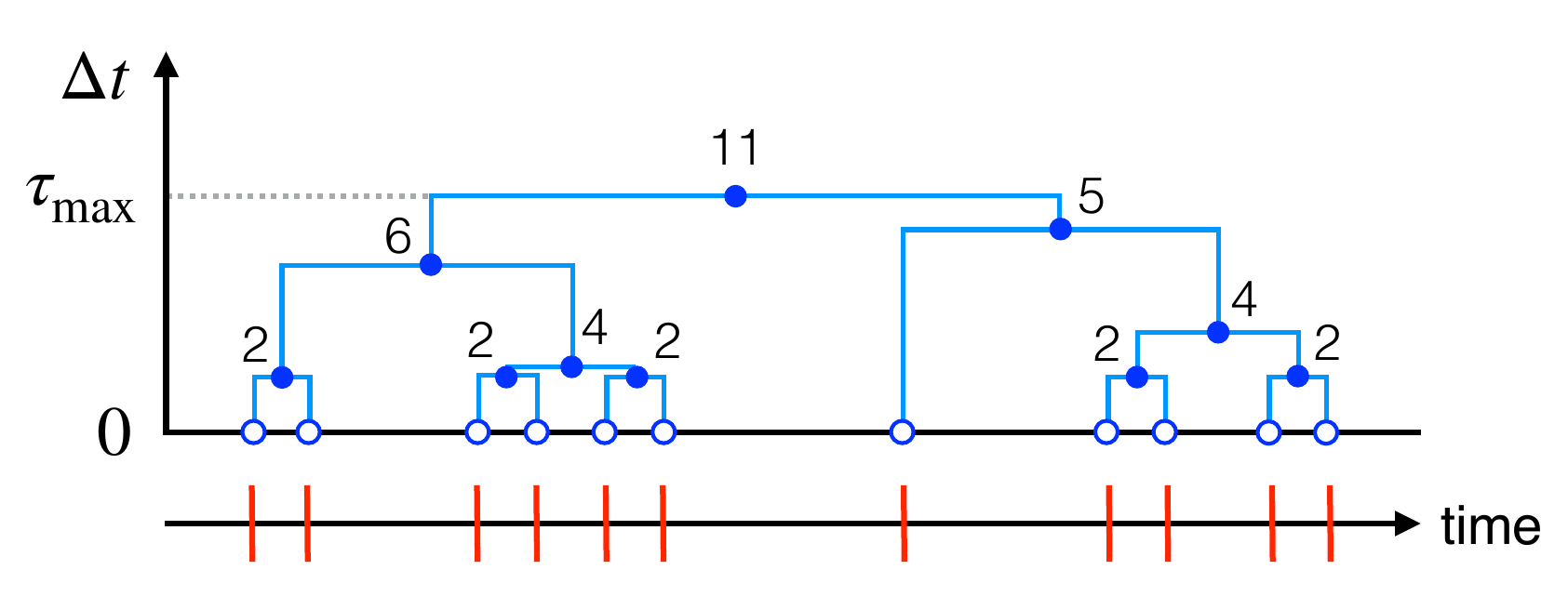}
\caption{Illustration of the burst-tree decomposition method. The lower panel shows time series where each red vertical line represents an event. In the upper panel, the vertical axis indicates the timescale $\Delta t$ for defining bursts; for $\Delta t<\tau_{\rm min}$, each event makes a burst of size one (empty circles). As $\Delta t$ increases, bursts are sequentially merged to form bigger ones (filled circles). When $\Delta t\geq \tau_{\rm max}$, all events belong to the single burst. The numbers next to the circles denote burst sizes.}
\label{fig:burst_tree_fig}
\end{figure}

The burst-tree decomposition method~\cite{Jo2020Bursttree} was proposed to fully reveal the hierarchical structure of bursts in the time series in an alternate form of a burst tree, as depicted in Fig.~\ref{fig:burst_tree_fig}. Let us consider an event sequence of $n$ events, denoted as $\mathcal{E} = \{t_0, \ldots, t_{n-1}\}$, where $t_i$ represents the timing of the $i$th event. The interevent time (IET) sequence, $\{\tau_1, \ldots, \tau_{n-1}\}$, is derived from the event sequence $\mathcal{E}$ using $\tau_i \equiv t_i - t_{i-1}$. We denote by $\tau_{\rm min}$ and $\tau_{\rm max}$ the minimum and maximum IETs in the IET sequence, respectively. For a given timescale $\Delta t$, events are clustered into bursts; IETs between two consecutive events within the same burst are smaller than or equal to $\Delta t$, while IETs between different bursts are greater than $\Delta t$. The number of events in a burst is called a burst size, and it is denoted by $b$. If $\Delta t$ is smaller than $\tau_{\rm min}$, each event makes its own burst of size one. Those bursts are depicted as empty circles in Fig.~\ref{fig:burst_tree_fig}. As $\Delta t$ continuously increases from $\tau_{\rm min}$, bursts of size one are merged with each other to become bigger bursts. Whenever $\Delta t$ exceeds an IET in the IET sequence, two successive bursts separated by the IET are merged, which is depicted as a filled circle in Fig.~\ref{fig:burst_tree_fig}. Finally, for $\Delta t\geq \tau_{\rm max}$, all events in the event sequence belong to the single burst. Such a merging pattern over the entire range of timescales can be visualized as a tree, thus it is called a burst tree. In Fig.~\ref{fig:burst_tree_fig}, empty circles and filled circles are called leaf nodes and internal nodes, respectively. 

The burst-tree decomposition method assumes only binary merging~\cite{Jo2020Bursttree}. Since each internal node means a merge of two bursts, it is also a parent node. A parent node indexed by $u$ has a left child node $v$ and a right child node $w$. Note that $v$ and $w$ can be either internal or leaf nodes. If their associated burst sizes are denoted by $b_u$, $b_v$, and $b_w$, respectively, one gets $b_u = b_v + b_w$. Since each merge occurs when $\Delta t$ equals to the IET separating left and right children nodes, the parent node $u$ is associated with the IET between the last event in the left child burst and the first event in the right child burst, denoted by $\hat \tau_{u}$; the set of $\hat \tau_{u}$ is identical to the IET distribution $P(\tau)$. Finally, the index $u$ is determined as the rank of the associated IET among all IETs, implying that $\hat \tau_i\geq \hat\tau_j$ for $i< j$ for all $i,j$. For example, the root node's index is $1$ as it is associated with $\tau_{\rm max}$, while the internal node whose associated IET is $\tau_{\rm min}$ has an index of $n-1$. Then, each internal node in the tree is described by the tuple $(u,v,w, \hat \tau_{u})$ and the burst tree $\mathcal{T}$ is represented by the collection of these tuples, i.e., $\{(u,v,w, \hat \tau_{u})\}$. The burst tree is then decomposed into the ordinal burst tree, denoted by $\mathcal{G}\equiv \{(u,v,w)\}$, and the IET distribution $P(\tau)$. It means that the same ordinal burst tree can be combined with an arbitrary IET distribution to generate event sequences with the same ordinal burs-tree structure but with different IET distributions. 

We remark that since the burst tree carries exactly the same information as the event sequence, one can reconstruct the time series from the burst tree. The timings are determined by using the formula $t_{i} = t_{i-1} +\hat \tau_{u(i)}$, where $i = 1,\ldots, n-1$. Here $u(i)$ indicates the $i$th visited node when traversing the burst tree in inorder. Note that the value of $t_0$ can be assigned either as the $t_0$ of the original data or as $0$ for convenience.

\begin{figure}[!t]
\centering
\includegraphics[width=0.9\columnwidth]{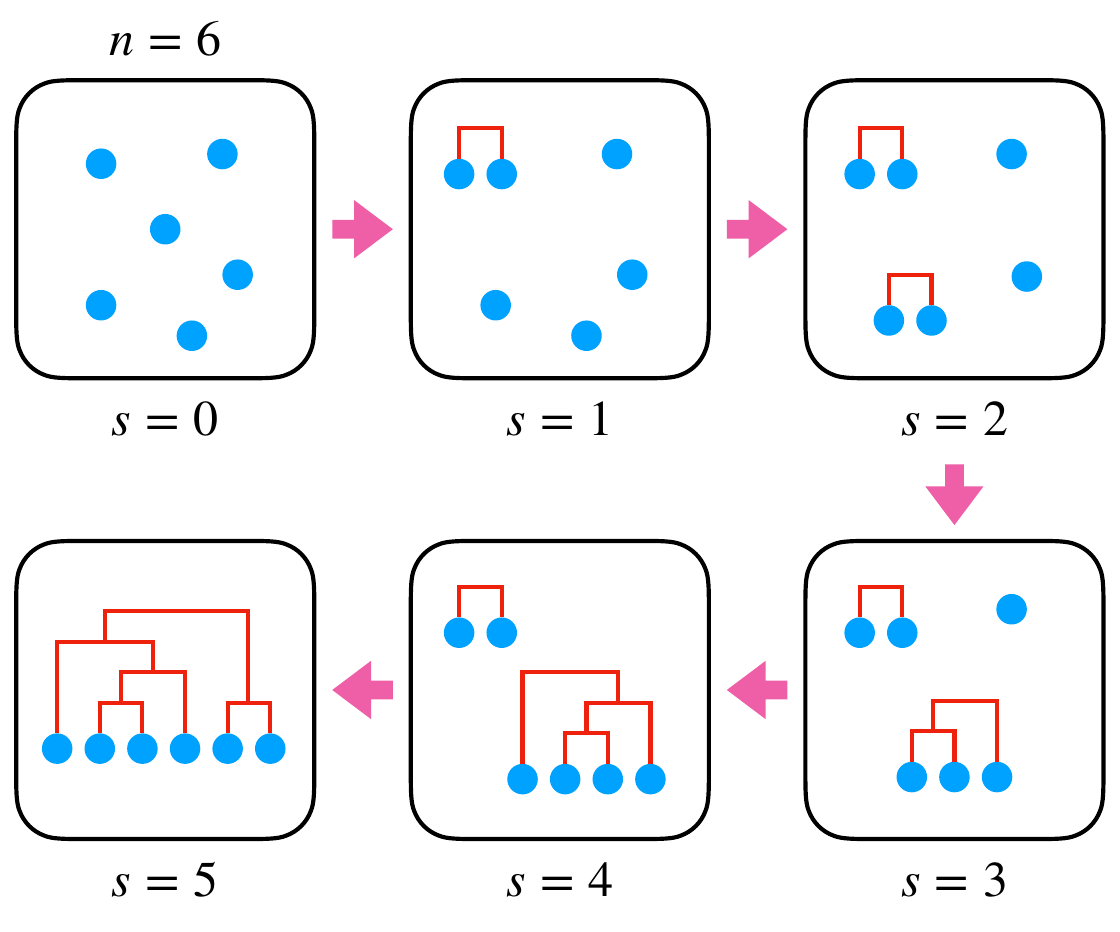}
\caption{Schematic diagram of the merging process for the ordinal burst tree with $n=6$.}
\label{fig:merging_process}
\end{figure}

\subsection{Burst-merging kernel for the ordinal burst tree}\label{subsec:kernel}

The ordinal burst tree $\mathcal{G}$ derived from the time series can be considered as a realization of the stochastic merging process driven by the burst-merging kernel $K_{bb'}$. For showing this, we make an analogy between the ordinal burst tree and the coagulation processes in statistical physics~\cite{Stockmayer1943Theory, Lushnikov1973Evolution, White1982Form, Hendriks1983Coagulation, Aldous1999Deterministic, Lee2001Survey, Leyvraz2003Scaling, Leyvraz2005Rigorous, Wattis2006Introduction, Birhanu2025Bursttree}. In the typical mean-field coagulation process, particles of unit mass are initially given. They are randomly chosen and merged to become clusters according to some selection rule or kernel. When only the binary merging is allowed, the kernel is written as a function of masses of two clusters, i.e., $K_{xy}$, meaning that the probability of choosing two clusters of masses $x$ and $y$ is proportional to $K_{xy}$. Then two chosen clusters of masses $x$ and $y$ are merged to form the cluster of mass $x+y$. As the coagulation is repeated, a broad range of cluster masses emerge as long as $K_{xy}$ is an increasing function of $x$ and/or $y$. One of the main interests in the coagulation process is to derive the time evolution of the cluster mass distribution.

Similarly, let us consider a merging process for the ordinal burst tree, as illustrated in Fig.~\ref{fig:merging_process}. We begin with $n$ events, equivalently, $n$ bursts of size one. Then we randomly choose two bursts using the burst-merging kernel $K_{bb'}$ and those chosen bursts are merged to make one burst of size $b+b'$. This merging process is repeated until all events belong to the single burst of size $n$. In contrast to the mean-field coagulation process, bursts are ordered in time. Thus, $K_{bb'}$ is considered being the kernel for choosing one burst of size $b$ as the earlier burst and the other burst of size $b'$ as the later burst. The earlier (later) burst corresponds to the left (right) child node of the parent node with burst size $b+b'$ in the ordinal burst tree. It implies that $K_{bb'}$ is not necessarily symmetric with respect to the exchange of $b$ and $b'$. 

To explicitly describe such a merging process, we introduce an auxiliary time step $s$ ranging from $0$ to $n-1$, which also equals to the number of merges; $s=0$ means the initial condition with $n$ bursts of size one (corresponding to $\Delta t<\tau_{\rm min}$), while $s=n-1$ does the final state with the single burst of size $n$ (corresponding to $\Delta t\geq \tau_{\rm max}$). At each time step $s$, we have exactly $n-s$ bursts and the distribution of their sizes is denoted by $Q_s(b)$. The $s$th merge is represented by a function $m_{sbb'}$, which is the number of merges of a burst of size $b$ and its successive burst of size $b'$ at the time step $s$. By definition of the auxiliary time step, one has $m_{sbb'}=1$ for only one pair of burst sizes $b$ and $b'$ at each $s$, otherwise $m_{sbb'}=0$, thus satisfying
\begin{align}
    m_s\equiv \sum_{b,b'=1}^n m_{sbb'}=1\ \text{for}\ s=1,\ldots,n-1.
    \label{eq:ms=1}
\end{align}
These quantities, i.e., $Q_s(b)$ and $m_{sbb'}$, will be used to estimate the burst-merging kernel in the next section.

\section{Burst-merging kernel estimation}\label{sec:estimation}

We first introduce the previous method for estimating the burst-merging kernel~\cite{Jo2020Bursttree}. Precisely, the burst-merging kernel has been estimated by the following method:
\begin{align}
    K_{bb'}=\frac{\sum_{s=1}^{n-1}m_{sbb'}}{\sum_{s=1}^{n-1} Q_{s-1}(b)Q_{s-1}(b')},
    \label{eq:K_bb'_2020}
\end{align}
which is based on the assumption that the expectation value of $m_{sbb'}$ must be proportional to the kernel $K_{bb'}$, multiplied by the probability of choosing bursts of sizes $b$ and $b'$, i.e., $Q_{s-1}(b)Q_{s-1}(b')$. Although this method has been successfully applied to several empirical datasets~\cite{Jo2020Bursttree}, it has not been proven to be rigorous, requiring us to devise a more statistically sound method.

\subsection{Maximum likelihood estimation method}\label{subsec:MLE}

Our aim is to estimate the burst-merging kernel $K_{bb'}$ from the given ordinal burst tree $\mathcal{G}$ in a similar manner to the maximum likelihood estimation method of the attachment kernel for growing networks~\cite{Pham2015PAFit}. Since $1\leq b\leq n$, we have a matrix of $n\times n$ elements for the kernel, denoted by $\textbf{K} =\{K_{bb'}\}_{b,b'=1,\ldots,n}$; note that $K_{bb'}$s for $b+b'>n$ make no physical sense. Let $G_s$ denote the snapshot of the merging process at the auxiliary time step $s$ (Fig.~\ref{fig:merging_process}). Then the likelihood of the ordinal burst tree data for the parameter matrix \textbf{K} is written as
\begin{align}
    L(\textbf{K})=P(\{G_s\}|\textbf{K})=\prod_{s=1}^{n-1} P(G_s|G_{s-1},\textbf{K})\cdot P(G_0),
\end{align}
where we have assumed that $G_{s}$ is conditioned only on $G_{s-1}$ not on $G_{s'}$ for $s'\leq s-2$. Here $P(G_s|G_{s-1},\textbf{K})$ is a multinomial distribution:
\begin{align}
    P(G_s|G_{s-1},\textbf{K})=\binom{m_s}{m_{s11},\ldots, m_{snn}}\prod_{b,b'=1}^n p_{sbb'}^{m_{sbb'}},
\end{align}
where
\begin{align}
    p_{sbb'}\equiv \frac{Q_{s-1}(b)Q_{s-1}(b')K_{bb'}}{\sum_{k,k'=1}^n Q_{s-1}(k)Q_{s-1}(k')K_{kk'}}.
    \label{eq:p_sbb'}
\end{align}
From Eq.~\eqref{eq:p_sbb'}, it is obvious that multiplying a constant to \textbf{K} does not affect $p_{sbb'}$, thus only the functional form of the kernel is relevant. We then derive the log-likelihood function by dropping all the terms not involving \textbf{K} as follows:
\begin{align}
    l(\textbf{K})&=\sum_{s=1}^{n-1}\sum_{b,b'=1}^n m_{sbb'}\ln K_{bb'}\notag\\
    &- \sum_{s=1}^{n-1}\ln \left[ \sum_{k,k'=1}^n Q_{s-1}(k)Q_{s-1}(k')K_{kk'}\right],
    \label{eq:l_K}
\end{align}
where for the second term on the right hand side, we have used Eq.~\eqref{eq:ms=1}. The derivative of the log-likelihood with respect to \textbf{K} is set to zero, i.e., $\partial l(\textbf{K})/\partial \textbf{K}=0$, to obtain
\begin{align}
    K_{bb'}=\dfrac{\sum_{s=1}^{n-1}m_{sbb'}}
    {\sum_{s=1}^{n-1} \dfrac{Q_{s-1}(b)Q_{s-1}(b')}{\sum_{k,k'=1}^n Q_{s-1}(k)Q_{s-1}(k')K_{kk'}}}.
    \label{eq:K_bb'_sol}
\end{align}
Note that Eq.~\eqref{eq:K_bb'_sol} is the self-consistent equation of \textbf{K} and nontrivial to derive its analytical solution. Thus, we estimate \textbf{K} by numerically iterating Eq.~\eqref{eq:K_bb'_sol}; we set the initial values of \textbf{K} as 
\begin{align}
    K^{(0)}_{bb'}=1\ \text{for}\ b,b'=1,\ldots,n.
\end{align}
Then, the $i$th iteration of \textbf{K}, i.e., $\textbf{K}^{(i)}$, for $i\geq 1$ is calculated using
\begin{align}
    K^{(i)}_{bb'}=\dfrac{\sum_{s=1}^{n-1}m_{sbb'}}
    {\sum_{s=1}^{n-1} \dfrac{Q_{s-1}(b)Q_{s-1}(b')}{\sum_{k,k'=1}^n Q_{s-1}(k)Q_{s-1}(k')K^{(i-1)}_{kk'}}}.
    \label{eq:K_bb'_iter}
\end{align}
The calculation of Eq.~\eqref{eq:K_bb'_iter} is repeated until the following convergence condition is satisfied:
\begin{align}
    \frac{|l(\textbf{K}^{(i)})-l(\textbf{K}^{(i-1)})|}{|l(\textbf{K}^{(i-1)})|+1}\leq \epsilon,
    \label{eq:convergence}
\end{align}
where $\epsilon$ is a convergence parameter having a sufficiently small positive value~\cite{Pham2015PAFit}. The implementation codes of our method are available in the GitHub repository~\cite{Birhanu2025Codes}.

Similarly to the previous work on the attachment kernel estimation for growing networks~\cite{Pham2015PAFit}, we have proved that the kernel \textbf{K} satisfying Eq.~\eqref{eq:K_bb'_sol} globally maximizes the log-likelihood function in Eq.~\eqref{eq:l_K} (see Appendix~\ref{append:proof1}), and that the log-likelihood function at $\textbf{K}^{(i)}$ increases with the iteration in Eq.~\eqref{eq:K_bb'_iter} (see Appendix~\ref{append:proof2}). In conclusion, our estimation method using the iteration in Eq.~\eqref{eq:K_bb'_iter} indeed converges to the optimal burst-merging kernel maximizing the likelihood. 

\begin{figure}[!t]
\centering
\includegraphics[width=\columnwidth]{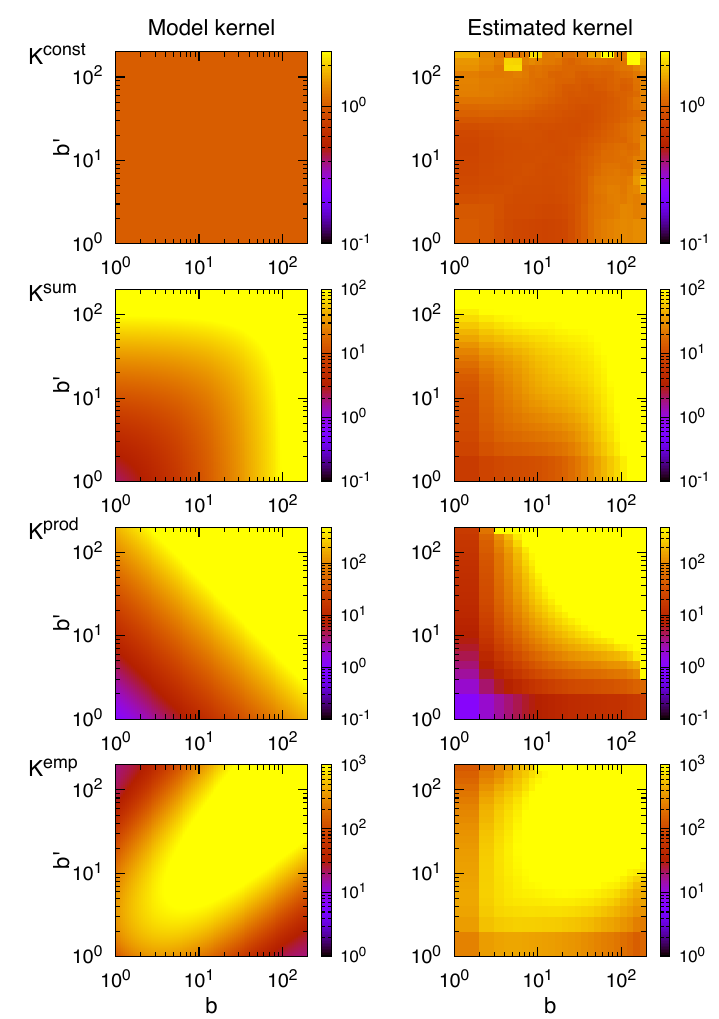}
\caption{Model kernels in Eqs.~\eqref{eq:K_const}--\eqref{eq:K_emp} (left) used to generate $100$ different time series with $10^5$ events for the IET distribution in Eq.~\eqref{eq:Ptau} with $\alpha=1.8$ and $\tau_{\rm c}=10^7$, and their corresponding estimated kernels (right) from the generated time series using our maximum likelihood estimation method with $\epsilon=10^{-4}$ in Eq.~\eqref{eq:convergence}. 
}
\label{fig:model}
\end{figure}

\subsection{Validation of the estimation method}\label{subsec:valid}

To demonstrate the validity of our estimation method, we generate time series using several model kernels, and estimate the kernels from the generated time series using our method to see whether the estimated kernels are in good agreement with model kernels. For this, we employ the following four burst-merging kernels:
\begin{align}
    &K_{bb'}^{\rm const}=1, \label{eq:K_const}\\
    &K_{bb'}^{\rm sum}=b+b', \label{eq:K_sum}\\
    &K_{bb'}^{\rm prod}=bb', \label{eq:K_prod}\\
    &K_{bb'}^{\rm emp} =\left[1 + c_1\ln (bb')\right] \left[1 + c_2e^{-c_3(\ln b - \ln b')^{2}}\right], \label{eq:K_emp}
\end{align} 
where $c_1=3$, $c_2=100$, and $c_3=1/4$ are used. We remark that all kernels in Eqs.~\eqref{eq:K_const}--\eqref{eq:K_emp} have been studied in Ref.~\cite{Birhanu2025Bursttree}. The first three, simplistic kernels enable to derive the analytical solutions of the burst-size distributions~\cite{Wattis2006Introduction}. The sum and product kernels are increasing with $b$ and $b'$, implying that bigger bursts are more likely to be chosen than smaller ones for the merge, making those bigger bursts even bigger. Thus, such a property has been called preferential merging. The last kernel $K_{bb'}^{\rm emp}$ in Eq.~\eqref{eq:K_emp} is based on the empirical results using Eq.~\eqref{eq:K_bb'_2020} for several datasets~\cite{Jo2020Bursttree}. It is increasing with $b$ and/or $b'$, while it is decreasing with the difference between $b$ and $b'$. It implies that bursts of similar sizes tend to be merged more often than those of different sizes. Thus, it is called assortative merging. All the model kernels are visualized for the range of $(b,b')\in [1,200]\times[1,200]$ in the left panels of Fig.~\ref{fig:model}. 

For each kernel mentioned above, we generate the ordinal burst tree for $n$ events as follows (see also Fig.~\ref{fig:merging_process}):
\begin{enumerate}
    \item At the initial time step $s=0$, each event makes its own burst of size one.
    \item At each time step $s$, two bursts of sizes $b$ and $b'$ are chosen at random with a probability proportional to $K_{bb'}$ [Eq.~\eqref{eq:p_sbb'}].
    \item Those two bursts are merged to make another burst of size $b+b'$, which becomes a parent node. Two merged bursts are assigned to left and right children nodes at random.
    \item Steps 2--3 are repeated until all events belong to the single burst of size $n$.
\end{enumerate}
Once the ordinal burst tree is generated, we randomly draw $n-1$ IETs from the power-law IET distribution as
\begin{align}
    P(\tau)=\frac{\tau^{-\alpha}}{\sum_{\tau=1}^{\tau_{\rm c}}\tau^{-\alpha}}\ \text{for}\ \tau=1,\ldots,\tau_{\rm c},
    \label{eq:Ptau}
\end{align}
where $\alpha$ is the power-law exponent and $\tau_{\rm c}$ is the upper bound. These IETs are sorted in a descending order to obtain $\{\hat\tau_1,\ldots,\hat\tau_{n-1} \}$, which are combined with the ordinal burst tree to generate the time series. That is, the timings of events are determined by using the formula $t_{i} = t_{i-1} +\hat \tau_{u(i)}$, where $i = 1,\ldots, n-1$. Here $u(i)$ indicates the $i$th visited node when traversing the ordinal burst tree in inorder. We set $t_0=0$ in all cases.

For the simulations, we generate $100$ different time series with $10^5$ events using each of burst-merging kernels in Eqs.~\eqref{eq:K_const}--\eqref{eq:K_emp} for the IET distribution with $\alpha=1.8$ and $\tau_{\rm c}=10^7$. Then we apply our method with $\epsilon=10^{-4}$ in Eq.~\eqref{eq:convergence} to estimate the burst-merging kernel for each of $100$ time series and take the average of them to obtain the estimated kernel $\hat{\textbf{K}}$. To compare the estimated kernel to the corresponding model kernel on the same scale, we normalize the estimated kernel as follows:
\begin{align}
    \hat{\textbf{K}'}= c\hat{\textbf{K}},\ c\equiv \frac{\sum_{b,b=1}^{100}K_{bb'}}{\sum_{b,b=1}^{100}\hat K_{bb'}},
\end{align}
where $K_{bb'}$ denotes the model kernel. When defining $c$, we have excluded the range of $b,b'>100$ as the values of $\hat{K}_{bb'}$ for that range show large fluctuations. We also logarithmically bin the normalized kernel by taking the average of $\hat{K}'_{bb'}$ for each bin, making the estimated kernel more stable. The results are shown in the right panels of Fig.~\ref{fig:model}, where we have limited the range of $(b,b')$ to $[1,200]\times[1,200]$. We find that the estimated kernels are in good agreement with the model kernels in all cases, validating our estimation method.

\begin{figure}[!t]
\centering
\includegraphics[width=\columnwidth]{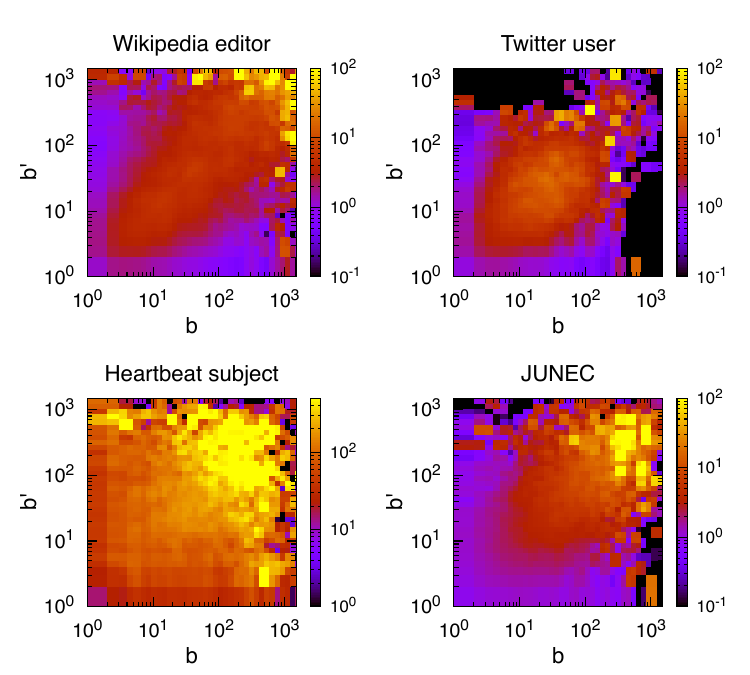}
\caption{Estimated kernels from four empirical time series described in the main text using the estimation method with $\epsilon=10^{-4}$ in Eq.~\eqref{eq:convergence} for the cases with Wikipedia editor and Twitter user, and $\epsilon=10^{-5}$ for the cases with heartbeat subject and JUNEC, respectively. Black squares imply no data in the bins of $(b,b')$.
}
\label{fig:dataset}
\end{figure}

\subsection{Application to empirical time series}\label{subsec:empirical}

We now apply our method to four empirical time series that have been analyzed in Ref.~\cite{Jo2020Bursttree}; they are (i) the edit sequence of $n\approx 1.1\times 10^6$ by the most active editor in the English Wikipedia~\cite{English}, (ii) the tweet sequence of $n\approx 8.7\times 10^4$ by the most active user of Twitter (currently X)~\cite{Yang2011Patterns}, (iii) the heartbeat time series of $n\approx 1.1\times 10^5$ measured in a healthy subject~\cite{PhysioBank}, and (iv) the earthquake sequence of $n\approx 2.0\times 10^5$ in the Japan University Network Earthquake Catalog (JUNEC)~\cite{Japan}. For the estimation, we have used $\epsilon=10^{-4}$ in Eq.~\eqref{eq:convergence} for the Wikipedia editor and the Twitter user, and $\epsilon=10^{-5}$ for the heartbeat subject and the JUNEC, respectively. The results shown in Fig.~\ref{fig:dataset} are overall similar to those reported in Ref.~\cite{Jo2020Bursttree}, but our current method must be more accurate than the previous estimation method in Eq.~\eqref{eq:K_bb'_2020}. In all cases, we find both preferential and assortative merging behaviors, providing insights into the hierarchical organization of bursts in time series.

\section{Conclusion}\label{sec:conclusion}

We have developed the maximum likelihood estimation method of the burst-merging kernel for characterizing the hierarchical structure of bursts in the time series. Thanks to the burst-tree decomposition method~\cite{Jo2020Bursttree}, any time series in the form of the event sequence can be exactly mapped to the burst tree, which fully reveals the hierarchical structure of bursts. The ordinal part of the burst tree can be summarized in terms of the burst-merging kernel dictating which bursts are merged together as the timescale defining bursts increases. Therefore, accurate estimation of the burst-merging kernel is of utmost importance to study the underlying mechanism behind the burst-tree structure. By employing the maximum likelihood estimation method, we could estimate the burst-merging kernels accurately, paving the way to study the underlying mechanism of the time series more rigorously. 

We remark that considering the similarity between the merging process for the time series and the coagulation processes~\cite{Birhanu2025Bursttree}, our method can be applied to the possible estimation of the merging kernels in the coagulation processes and other related physical processes. In the context of network science, if new links are added to the network depending on the degrees of both nodes to be connected, such a mechanism might be understood by the similar method as ours; one can estimate the kernel for choosing two existing nodes for making a connection between those nodes.


\begin{acknowledgments}
T.B. and H.-H.J. acknowledge financial support by the National Research Foundation of Korea (NRF) grant funded by the Korea government (MSIT) (No. 2022R1A2C1007358).
\end{acknowledgments}

\appendix

\section{Proof of the global maximization of the log-likelihood function}\label{append:proof1}

To prove that the kernel satisfying Eq.~\eqref{eq:K_bb'_sol} globally maximizes the log-likelihood function in Eq.~\eqref{eq:l_K}, we introduce a new matrix of $n\times n$ elements for $\textbf{g}=\{g_{bb'}\}_{b,b'=1,\ldots,n}$ with the definition:
\begin{align}
    g_{bb'}\equiv \ln K_{bb'}.
\end{align}
The log-likelihood function is written as
\begin{align}
    l(\textbf{g})&=\sum_{s=1}^{n-1}\sum_{b,b'=1}^n m_{sbb'}g_{bb'}\notag\\
    &- \sum_{s=1}^{n-1}\ln \left[ \sum_{k,k'=1}^n Q_{s-1}(k)Q_{s-1}(k')e^{g_{kk'}}\right].
    \label{eq:l_g}
\end{align}
For simplicity, we introduce the notation as $f_{sbb'}\equiv Q_{s-1}(b)Q_{s-1}(b')e^{g_{bb'}}$; note that $f_{sbb'}\geq 0$. We then take the second derivative of $l(\textbf{g})$ with respect to $g_{bb'}$ as follows:
\begin{align}
    \frac{\partial^2 l(\textbf{g})}{\partial g_{bb'}^2}
    =-\sum_{s=1}^{n-1} \frac{f_{sbb'}(\sum_{k,k'=1}^n f_{skk'}-f_{sbb'})}
    {(\sum_{k,k'=1}^n f_{skk'})^2}\leq 0,
\end{align}
which holds for any $b,b'$. It implies that $l(\textbf{g})$ is a concave function of \textbf{g}, hence it is a concave function of \textbf{K} too. Thus, the kernel obtained from the condition $\partial l(\textbf{K})/\partial \textbf{K}=0$ makes $l(\textbf{K})$ globally maximized.

\section{Proof of the increasing log-likelihood function with the iteration in Eq.~\eqref{eq:K_bb'_iter}}\label{append:proof2}

To prove that the iteration of the kernel in Eq.~\eqref{eq:K_bb'_iter} increases the log-likelihood function at the iterated kernel, we adopt the minorize-maximize (MM) algorithm~\cite{Hunter2004Tutorial}. We first construct the minorizing function for the $i$th iterated $\textbf{K}^{(i)}$ in Eq.~\eqref{eq:K_bb'_iter}, which is denoted by $Q_i(\textbf{K})$. The minorizing function $Q_i(\textbf{K})$ should satisfy the following two conditions:
\begin{align}
\label{eq:l_K1}
    &l(\textbf{K}) \geq Q_{i}(\textbf{K}), \\
    &l(\textbf{K}^{(i)}) = Q_{i}(\textbf{K}^{(i)}).
    \label{eq:l_K2}
\end{align}
We minorize only the second term of $l(\textbf{K})$ in Eq.~\eqref{eq:l_K} by using the inequality $-\ln x\geq -\ln y-x/y+1$ that holds for any $x,y>0$, and then by plugging
\begin{align}
    &x = \sum_{kk' = 1}^{n} Q_{s-1}(k)Q_{s-1}(k')K_{kk'}, \notag\\
    &y = \sum_{kk' = 1}^{n} Q_{s-1}(k)Q_{s-1}(k')K_{kk'}^{(i)}.
\end{align}
We get the minorizing function as follows:
\begin{align}
    Q_i(\textbf{K}) &=\sum_{s=1}^{n-1}\sum_{b,b'=1}^n m_{sbb'}\ln K_{bb'} \notag\\
    &- \sum_{s=1}^{n-1}\ln \left[ \sum_{k,k'=1}^n Q_{s-1}(k)Q_{s-1}(k')K_{kk'}^{(i)}\right] \notag\\
    &-\sum_{s=1}^{n-1} \frac{\sum_{k,k'=1}^n Q_{s-1}(k)Q_{s-1}(k')K_{kk'}}{ \sum_{k,k'=1}^n Q_{s-1}(k)Q_{s-1}(k')K_{kk'}^{(i)}} + n-1.
    \label{eq:minorize}
\end{align}
Now we find the kernel $\textbf{K}^*$ maximizing the minorizing function $Q_i(\textbf{K})$ by solving
\begin{align}
    \frac{\partial Q_i(\textbf{K})}{\partial \textbf{K}}\bigg|_{\textbf{K}=\textbf{K}^*}=0.
\end{align}
Then it turns out that $\textbf{K}^*=\textbf{K}^{(i+1)}$ by the iteration formula in Eq.~\eqref{eq:K_bb'_iter}, indicating that
\begin{align}
    Q_i(\textbf{K}^{(i)})\leq Q_i(\textbf{K}^{(i+1)}).
    \label{eq:QiQi}
\end{align}
Combining Eqs.~\eqref{eq:l_K1} and~\eqref{eq:l_K2} with the inequality in Eq.~\eqref{eq:QiQi}, we finally conclude that
\begin{align}
    l(\textbf{K}^{(i)})=Q_i(\textbf{K}^{(i)})\leq Q_i(\textbf{K}^{(i+1)})\leq l(\textbf{K}^{(i+1)}).
\end{align}
In other words, the log-likelihood function increases with the number of iterations in Eq.~\eqref{eq:K_bb'_iter}.

%

\end{document}